\documentclass[preprint,showkeys, preprintnumbers,amssymb]{revtex4}

\usepackage{graphicx}
\graphicspath{{figs/}}
\begin{document}
\title{Isotopic effects on the thermal conductivity of graphene nanoribbons: localization mechanism}
\author{Jin-Wu~Jiang}
\thanks{Electronic mail: phyjj@nus.edu.sg}
	\affiliation{Department of Physics and Center for Computational Science and Engineering,
 		     National University of Singapore, Singapore 117542, Republic of Singapore }
\author{Jinghua Lan}
        \affiliation{Institute of High Performance Computing, 1 Fusionopolis Way, \#16-16 Connexis, Singapore 138632, Republic of Singapore}
\author{Jian-Sheng~Wang}
	\affiliation{Department of Physics and Center for Computational Science and Engineering,
     		     National University of Singapore, Singapore 117542, Republic of Singapore }
        \affiliation{Institute of High Performance Computing, 1 Fusionopolis Way, \#16-16 Connexis, Singapore 138632, Republic of Singapore}
\author{Baowen~Li}
        \affiliation{Department of Physics and Center for Computational Science and Engineering,
                     National University of Singapore, Singapore 117542, Republic of Singapore }
        \affiliation{NUS Graduate School for Integrative Sciences and Engineering,
                     Singapore 117456, Republic of Singapore}
\date{\today}
\begin{abstract}
Thermal conductivity of graphene nanoribbons (GNR) with length 106~{\AA} and width 4.92~{\AA} after isotopic doping is investigated by molecular dynamics with quantum correction. Two interesting phenomena are found: (1) isotopic doping reduces thermal conductivity effectively in low doping region, and the reduction slows down in high doping region; (2) thermal conductivity increases with increasing temperature in both pure and doped GNR; but the increasing behavior is much more slowly in the doped GNR than that in pure ones. Further studies reveal that the physics of these two phenomena is related to the localized phonon modes, whose number increases quickly (slowly) with increasing isotopic doping in low (high) isotopic doping region.
\end{abstract}

\maketitle

\section{introduction}
The thermal conductivity is a interesting property of graphene. Recent experiments have measured the thermal conductivity to be an extremely high value.\cite{Balandin, Ghosh} Thermal conductivity
can be theoretically studied by quantum molecular dynamics (MD)\cite{Wang, WangJS2009}, classical MD\cite{Schelling, Maruyama, Maiti, Zhang, Wangsc, Lukes, Wu}, or Klemens perturbation theory approach\cite{NikaAPL}. For the 
classical MD, the temperature should be corrected to account for the quantum effect on the thermal conductivity. This quantum correction 
has been shown to be important below room temperature for nanotube\cite{Lukes} and nanowires\cite{Wangsc}.
Meanwhile, it has been demonstrated that the umklapp, defects and edges scattering have important effect on the thermal conductance in single layer graphene.\cite{Nika} Isotopic doping also has very important effect on thermal transport. For nanotube, both experimental\cite{Chang} and theoretical\cite{Zhang} results show that the isotopic doping can reduce the thermal conductivity remarkably.

In this paper we investigate the isotopic doping effect on the thermal conductivity in graphene nanoribbons (GNR). The thermal conductivity of GNR is studied by classical MD with quantum correction. Our studies display two phenomena: 
(1) in low isotopic doping region, the thermal conductivity decreases rapidly with increasing doping, and 10$\%$ doping can yield 50$\%$ reduction of the value of thermal conductivity; while in the high isotopic doping region, the thermal conductivity decreases slowly with further increasing doping.
(2) thermal conductivity of both pure GNR and isotopic doped GNR increases with increasing temperature, but the increasing behavior in doped GNR is slower than that in pure ones.

The underlying mechanism for these phenomena is the localization of phonon modes. (1) For the first phenomenon, we find that a single isotopic doping center can localize three phonon modes with the frequency as 634.4, 1373.3 and 1629.9 cm$^{-1}$. Due to their localizing property, these modes are unfavorable for thermal transport. So generally the thermal conductivity will decrease after isotopic doping. We further show that the number of localized modes increases linearly with increasing doping in the low doping region, while this number increases very slowly with increasing doping in the high doping region. This mechanism results in the first phenomenon. (2) For the second phenomenon, at higher temperature more phonon modes will be excited and the quantum correction is less important. These two positive factors facilitate thermal transport. So generally the thermal conductivity will increase with increasing temperature. However, in doped GNR, more localized modes will also be excited at higher temperature, leading to negative effect on thermal transport. This negative factor competes with the former two positive factors, slowing down the increasing behavior in isotopic doping GNR. It should be noted that the increase of the thermal conductivity for graphene nanoribbon is true for nanoribbons where the transport is dominated by the edge effects. However, in large flakes of graphene (larger than the phonon mean free path), the Umklapp scattering will lead to decreasing thermal conductivity with increasing temperature.\cite{Nika}

The rest of this paper is organized as follows. In Sec. II, thermal conductivity of graphene is calculated by MD, and results are interpreted by localization theory. Sec. III is the conclusion.

\section{molecular dynamics simulation}
\subsection{simulation details}
The thermal conductivity is calculated through the direct MD method with the second generation Brenner inter-atomic potential applied\cite{Brenner}. We apply periodic boundary condition in the width direction and fixed boundary condition in the length direction. The temperature for the left and right part is controlled to be constant value $T_{L}$ and $T_{R}$ by the No$\acute{s}$e-Hoover (NH) heat baths.\cite{Nose, Hoover} The dynamical equations for the NH heat bath are:
\begin{eqnarray}
\frac{dr_{i}}{dt} & = & \frac{p_{i}}{m_{i}},\nonumber\\
\frac{dp_{i}}{dt} & = & -\frac{\partial V}{\partial r_{i}}-\eta_{i} p_{i},\\
\frac{d\eta_{i}}{dt} & = & \left(\sum_{j}\frac{p_{j}^{2}}{m_{j}}-gk_{B}T\right)/Q,\nonumber\\\nonumber
Q & = & gk_{B}T\tau^{2},
\label{eq_nh_dynamics}
\end{eqnarray}
where $\eta_{i}=0$ if atom $i$ is not in the heat bath, and $g$ is the number of degrees of freedom for atoms in the heat bath. $\tau$ is the relaxation time of the heat baths and 
kept as 0.4 ps in our 
calculation. From the above dynamical equations, we can derive the exchanged energy between the heat bath and the
system:
\begin{eqnarray}
E_{nh} & = & -\int_{t_{0}}^{t_{0}+\Delta t}\sum_{i}\eta_{i}\frac{p_{i}^{2}}{m_{i}}dt,
\end{eqnarray}
where the summation of $i$ is taken over atoms in the heat bath. So the heat current is\cite{Poetzsch}:
\begin{eqnarray}
J & = & \frac{1}{s}\frac{E_{nh}}{\Delta t}.
\end{eqnarray}
In the steady state, the magnitude of heat current from the left and right heat baths should be equal: $J_{L}=-J_{R}$. So we can use $J=(J_{L}-J_{R})/2$ to 
calculate the heat current of the GNR. The difference between left and right heat current, i.e $dJ=|J_{L}|-|J_{R}|$, is
 used to determine whether the system has achieved steady state or not. Due to the numerical error, the value of $dJ$ in the steady state
is not strictly zero. This nonzero value is used to estimate the error bar 
for the value of thermal conductivity results throughout this paper.

We use 8.5 ns for the system to relax and do average with another 8.5 ns. Once the 
temperature gradient is obtained from this temperature profile, the thermal conductivity is then calculated by the Fourier law:
\begin{eqnarray}
J=-\kappa\frac{dT}{dx}.
\label{eq_Fourier}
\end{eqnarray}

\subsection{quantum correction}
\begin{figure}
  \begin{center}
    \scalebox{1.4}[1.4]{\includegraphics[width=7cm]{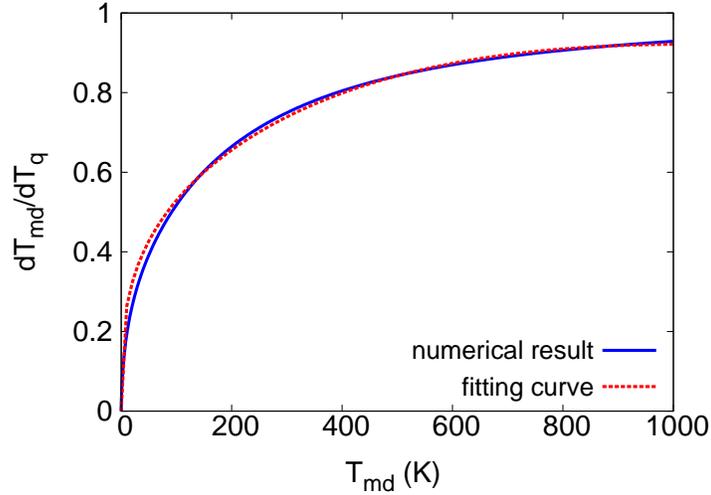}}
  \end{center}
  \caption{Quantum correction to thermal conductivity of graphene v.s MD temperature.}
  \label{fig_Tc_Tq}
\end{figure}

Since the Debye temperature $T_{D}$ of graphene is very high (above 1000~K),\cite{Benedict, Tohei, Falkovsky} it is necessary to consider 
the quantum effect for the thermal conductivity. We do quantum correction for the results of the classical MD 
in the following ways. Firstly, the quantum temperature $T_{q}$ is calculated by equaling the total ensemble 
energy to half of the total phonon energy:\cite{Maiti}
\begin{eqnarray}
\frac{3}{2}Nk_{B}T_{\rm md}=\frac{1}{2}\int_{0}^{\infty}g(\omega)n(\omega,T_{q})\hbar\omega d\omega,
\label{eq_Tc_Tq}
\end{eqnarray}
where $T_{md}$ is the MD simulation temperature. $g(\omega)$ is the density of states (DOS). We calculate $g(\omega)$ from the Brenner empirical potential implemented in the 
`General Utility Lattice Program' (GULP)\cite{Gale}. $n(\omega,T_{q})$ is the distribution function:
\begin{eqnarray}
n(\omega,T_{q})=\frac{1}{exp(\hbar\omega/k_{B}T_{q})-1}.
\end{eqnarray}
The thermal conductivity with quantum correction can be obtained from the classical MD results through following relationship\cite{Maiti}
\begin{eqnarray}
\kappa_{q}(T_{q})=\kappa_{\rm md}(T_{\rm md})\frac{\partial T_{\rm md}}{\partial T_{q}},
\end{eqnarray}
where $\kappa_{\rm md}$ is the MD result at temperature $T_{\rm md}$ and $\kappa_{q}$ is result with quantum correction. The correction factor $(\frac{\partial T_{md}}{\partial T_{q}})$ is shown in Fig.~\ref{fig_Tc_Tq}. It shows that in $T_{md}<300$K region, this quantum correction is more important, while the correction is small in the high temperature region. This tendency is qualitatively consistent with the previous results on carbon nanotubes\cite{Lukes} and silicon nanowires\cite{Wangsc}.

\subsection{results and discussion}
At $T_{\rm md}$=300 K, the value of the thermal conductivity of pure GNR in our calculation is about 61 W/(mK). The GNR has 400 carbon atoms, with length 106~{\AA} and width 4.92~{\AA}. This value
is comparable with that of the carbon nanotube with similar length\cite{Lukes}. These small absolute values of the thermal conductivity obtained in theoretical works are the results of the very small size of the examined graphene nanoribbon compared with the experimental graphene flakes. In the experiment, if we take the trench width as the length of graphene, the maximum thermal conductivity for a 10.6 nm long graphene is 5000/2000*10.6= 26.5 W/mK, which is in the same order of the present work but still quite different. This difference is probably results from the rough quantum correction.\cite{WangJS2009, Turney} Since our main focus is to study the isotopic doping effect on thermal conductivity, we accept this quantum correction for all calculations in present paper. More accurate quantum effect can be considered by other more precise method, such as the nonequilibrium Green's function method (see Ref.~\onlinecite{WangJS2008} for review).

\begin{figure}
  \begin{center}
    \scalebox{1.3}[1.3]{\includegraphics[width=7cm]{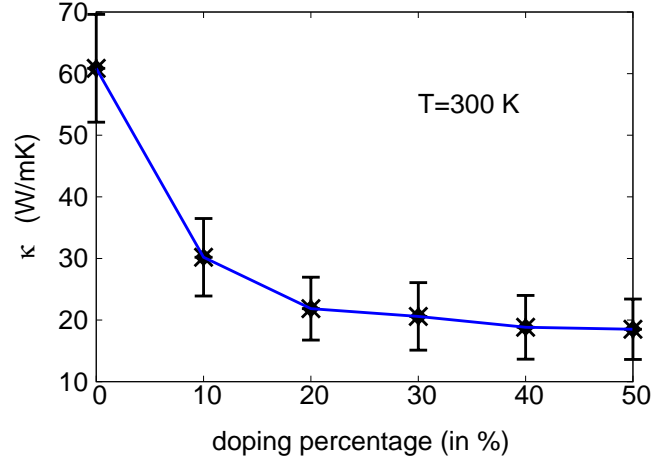}}
  \end{center}
  \caption{Thermal conductivity with quantum correction v.s isotope percentage at 300 K for GNR with length 106~{\AA} and width 4.92~{\AA}. Line is a guide to the eye.}
  \label{fig_isotope}
\end{figure}

Fig.~\ref{fig_isotope} shows the isotopic doping effect on the thermal conductivity. The carbon isotope $^{14}$C is added
by randomly substituting $^{12}$C in pure $^{12}$C GNR system, and the value of the thermal conductivity in the doped GNR is obtained by averaging over ten randomly doping processes.
It shows that the thermal conductivity decreases
rapidly before 10$\%$ doping, and decreases slowly after this doping percentage. With only 10$\%$ isotopic doping, the 
thermal conductivity is reduced up to 50$\%$, which is comparable with the results in nanotubes.\cite{Chang, Zhang}

In order to understand the underlying mechanism, it is essential to study localized phonon modes around the isotopic doping center. We apply the GULP to obtain the full force constant matrix of the GNR under the same boundary condition as used in the MD simulation. Then the force constant matrix is diagonalized to obtain the full phonon dispersion and eigen vector of the system. The localization property of each phonon mode can be analyzed from its corresponding eigen vector. If the component of one atom in the eigen vector is very large and decreases exponentially to its neighboring atoms, then this is the localized phonon mode around this particular atom.

\begin{figure}[htpb]
  \begin{center}
    \scalebox{1.0}[1.0]{\includegraphics[width=7cm]{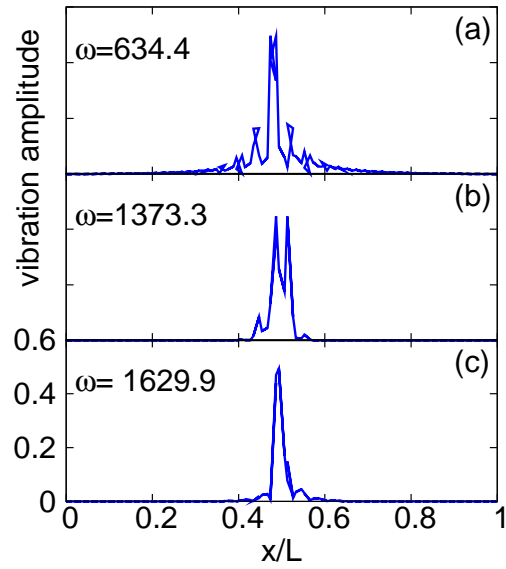}}
  \end{center}
  \caption{(Color online) Normalized vibration amplitudes vs. reduced $x$ for three phonon modes localized around one isotopic doping center in GNR with length 106~{\AA} and width 4.92~{\AA}. The frequency $\omega$ for each mode is in cm$^{-1}$.}
  \label{fig_iso_25_2_1}
\end{figure}

Fig.~\ref{fig_iso_25_2_1} shows the phonon modes localized around one $^{14}$C doping center. There are three localized modes around this $^{14}$C doping atom, with frequencies as 634.4, 1373.3 and 1629.9 cm$^{-1}$. Due to their localizing property, these localized modes have negative effect on the thermal conductivity. The frequencies of the first two localized mode are not very high. They can be excited in room temperature region. So these localized modes can reduce the thermal conductance effectively. 

When two $^{14}$C doping centers occur together in the GNR, there will be two different cases.

\begin{figure}[htpb]
  \begin{center}
    \scalebox{1.0}[1.0]{\includegraphics[width=7cm]{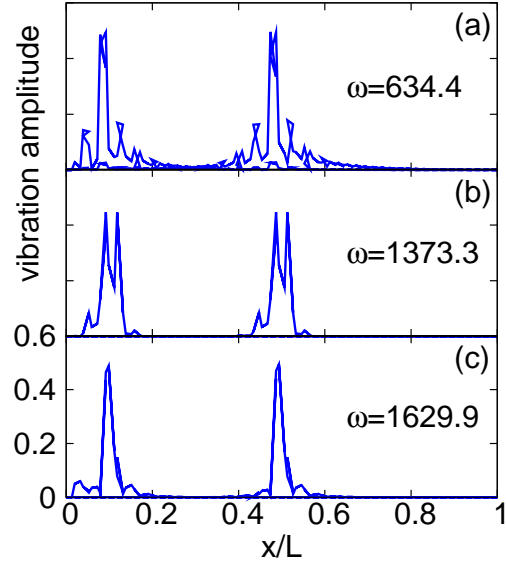}}
  \end{center}
  \caption{(Color online) Normalized vibration amplitudes vs. reduced $x$ of three localized modes in GNR with two isotopic doping centers, which are far from each other. All modes are degenerate. The frequency $\omega$ for each mode is in cm$^{-1}$.}
  \label{fig_iso_25_2_2_large}
\end{figure}

(1). In the first case, these two $^{14}$C doping atoms are far from each other. Fig.~\ref{fig_iso_25_2_2_large} shows that the frequencies of these three localized modes do not change. Since these two doping centers are far from each other, they can not `see' each other and these modes can be localized around either of these two $^{14}$C atoms, leading to degeneracy of all three modes. This degeneracy means that there are two phonon modes which have the same frequency. However, from the eigen vector of them we can see that they localized around different doping atoms. So in this case the number of localized modes will increase quickly (linearly) with increasing doping atoms.

\begin{figure}[htpb]
  \begin{center}
    \scalebox{1.0}[1.0]{\includegraphics[width=7cm]{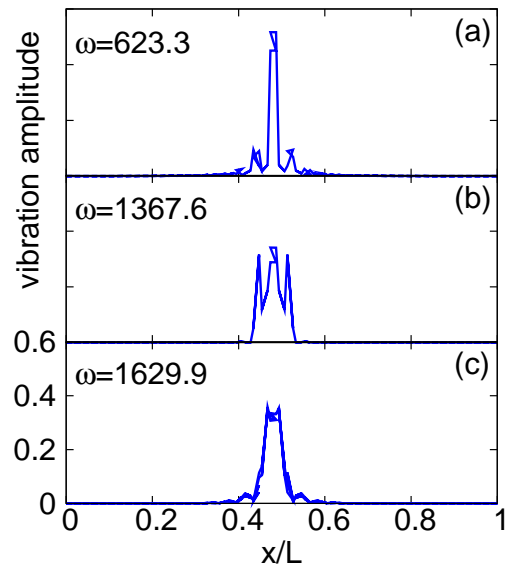}}
  \end{center}
  \caption{(Color online) Normalized vibration amplitudes vs. reduced $x$ of three edge modes in GNR with two isotopic doping centers, which are close with each other. Modes in (a) and (b) are non-degenerate, while the mode (c) is degenerate. The frequency $\omega$ for each mode is in cm$^{-1}$.}
  \label{fig_iso_25_2_2_small}
\end{figure}

(2). In the second case, these two $^{14}$C atoms are very close with each other. Fig.~\ref{fig_iso_25_2_2_small} indicates that these two $^{14}$C atoms act more like a single C-C molecule doping center. Their effect is to slightly change the frequency of the first two modes, but keep them non-degenerate. While the third mode becomes degenerate without changing frequency. So in this second case the number of localized modes increases very slowly with increasing doping atoms. We stress that the number of the first two low frequency localized modes does not increase. These two modes are more important than the third high frequency localized mode to reduce thermal conductivity.

We have also studied localized modes for more than two isotopic doping centers. The situation is similar as two doping centers described above. In the first case where isotopic doping centers are far from each other, the number of localized modes increases linearly. While in the second case, this number increases slowly.

We can now explain the isotopic doping effect on the thermal conductivity in Fig.~\ref{fig_isotope}. In the low doping region ($<10\%$), most isotopic doping atoms are far from each other, which is the first case mentioned above. So the number of localized modes increases rapidly with increasing doping. As a result, the thermal conductivity decreases quickly. In the high doping region most $^{14}$C atoms are close with each other. This is in the second case and the number of localized modes increases slowly with further increasing doping. Especially, the number of low frequency localized modes, which are more important in the reduction of thermal condctivity, does not increase too much. Consequently, the decreasing behavior of thermal conductivity with increasing doping slows down.

\begin{figure}
  \begin{center}
    \scalebox{1.3}[1.3]{\includegraphics[width=7cm]{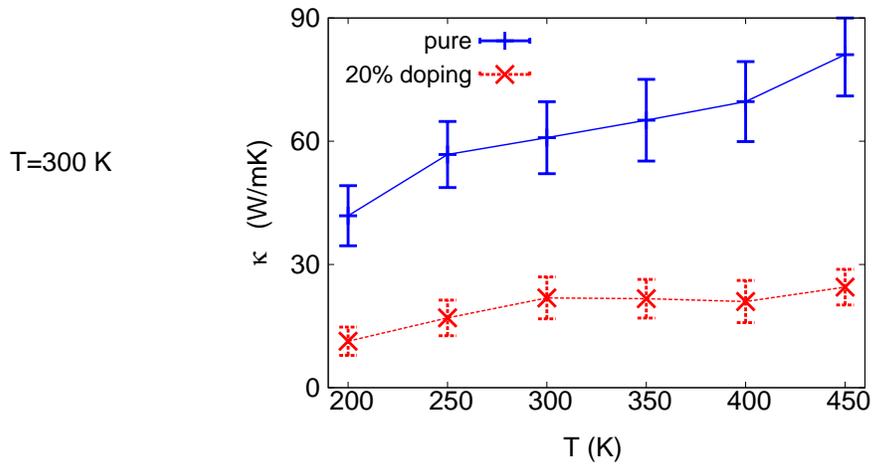}}
  \end{center}
  \caption{(Color online) The dependence of thermal conductivity on temperature for pure and 20$\%$ doped GNR.}
  \label{fig_tem}
\end{figure}

The temperature dependence for the thermal conductivity of pure and 20$\%$ doped GNR is shown in Fig.~\ref{fig_tem}. The common phenomenon in pure and doped GNR is that the thermal conductivity increases with increasing temperature. There are two reasons for this phenomenon. (1). At higher temperature, more high frequency phonon modes will be excited. which is beneficial for thermal transport. (2). As shown in Fig.~\ref{fig_Tc_Tq}, the quantum correction is more serious in lower temperature, which reduces the value of the thermal conductivity. The difference between these two system is that the thermal conductivity in the doped GNR increases much more slowly with increasing temperature than that in the pure GNR. This is because localized modes in the doped GNR are more excited at higher temperature. Thus slow down the increasing behavior of thermal conductivity with increasing temperature. We stress that the increase of thermal conductivity is only true in the graphene nanoribbon where the edge effect is important. In large flakes of graphene, the Umklapp scattering dominates the phonon transport, and leads to decreasing thermal conductivity with the increase of temperature.

\section{conclusion}
In conclusion, we study the thermal conductivity of GNR by means of molecular dynamics with quantum correction. It shows that the isotopic doping can reduce thermal conductivity remarkably by localizing phonon modes. Since the number of localized modes increase linearly with increasing doping in the low doping region, the reduction of thermal conductivity is distinct in the low isotopic doping region. But in the high doping region, the number of localized modes increase slowly with increasing doping, thus thermal conductivity decreases slowly with further increasing doping. It also shows that the thermal conductivity will increase with increasing temperature and isotopic doping can slow down this behavior by localizing phonon modes.

\section{Acknowledgements}
JJW acknowledges Jie Chen for helpful discussion on the quantum correction. The work is supported in part by a Faculty Research Grant of R-144-000-257-112 of NUS, and No. R-144-000-173-101/112 of NUS, and Grant No. R-144-000-203-112 from Ministry of Education of Republic of Singapore, and Grant No. R-144-000-222-646 from NUS.


\end{document}